\documentclass[12pt]{iopart}

\usepackage{graphicx}
\usepackage{amsfonts}
\usepackage{amssymb}
\usepackage{url}
\usepackage{epstopdf}
\usepackage{color}
\usepackage{bm}
\usepackage[colorlinks=true,linkcolor=blue,citecolor=red]{hyperref}%
\usepackage{cite}

\def\L{\mathcal L}
\def\N{\mathcal N}
\def\var{\mathrm{Var}}

\def\Ei{{\rm Ei}}

\def\x{\bm{x}}
\def\X{\bm{X}}

\def\ve{\varepsilon}

\def\Re{{\rm Re}}
\def\Im{{\rm Im}}

\def\pa{\partial\Omega}

\def\E{{\mathbb E}}

\def\P{{\mathbb P}}
\def\R{{\mathbb R}}
\def\Z{{\mathbb Z}}
\def\T{{\mathcal T}}
\def\L{{\mathcal L}}

\def\M{{\mathcal M}}

\def\erf{\mathrm{erf}}
\def\erfc{\mathrm{erfc}}

\def\var{\mathrm{var}}

\begin{document}

\title[Statistics of boundary encounters...]{Statistics of boundary encounters by a particle diffusing outside a compact planar domain}

\author{Denis~S.~Grebenkov}
 \ead{denis.grebenkov@polytechnique.edu}
% \email{denis.grebenkov@polytechnique.edu}
%\affiliation{
\address{
Laboratoire de Physique de la Mati\`{e}re Condens\'{e}e (UMR 7643), \\ 
CNRS -- Ecole Polytechnique, IP Paris, 91128 Palaiseau, France}

\date{\today}

\begin{abstract}
We consider a particle diffusing outside a compact planar set and
investigate its boundary local time $\ell_t$, i.e., the rescaled
number of encounters between the particle and the boundary up to time
$t$.  In the case of a disk, this is also the (rescaled) number of
encounters of two diffusing circular particles in the plane.  For that
case, we derive explicit integral representations for the probability
density of the boundary local time $\ell_t$ and for the probability
density of the first-crossing time of a given threshold by $\ell_t$.
The latter density is shown to exhibit a very slow long-time decay due
to extremely long diffusive excursions between encounters.  We briefly
discuss some practical consequences of this behavior for applications
in chemical physics and biology.
\end{abstract}


%02.50.-r       (Probability theory, stochastic processes, and statistics)
%05.40.-a 	Fluctuation phenomena, random processes, noise, and Brownian motion
%02.70.Rr       (General statistical methods)
%05.10.Gg 	Stochastic analysis methods (Fokker-Planck, Langevin, etc.) 

%02.50.Ey 	Stochastic processes  (Probability theory, stochastic processes, and statistics)


%\keywords{Boundary local time, Reflected Brownian motion, Diffusion-influenced reactions, 
%Surface reactivity, Robin boundary condition, Heterogeneous catalysis}

\submitto{\JPA}

\maketitle

\section{Introduction}
\label{sec:intro}

\begin{figure}
\begin{center}
\includegraphics[width=50mm]{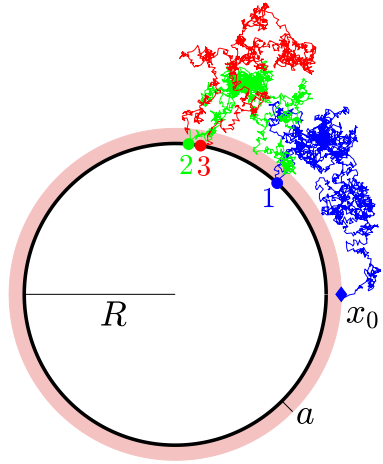} % disk_traj2_.eps}
\end{center}
\caption{
A simulated trajectory of reflected Brownian motion in the exterior of
a disk of radius $R$: $\Omega = \{ \x\in\R^2~:~|\x|>R\}$.  Pink region
denotes a thin boundary layer $\Gamma_a$ of width $a$ near the
boundary: $\Gamma = \pa$.  The particle is released in a close
vicinity of the boundary (blue diamond) and diffuses until the first
encounter with that boundary (filled circle enumerated by $1$).  From
that point, the particle is released a distance $a$ above the circle
and resumes its motion until the next encounter, and so on.  Such a
regularization with small $a> 0$ allows one to define the number of
encounters $\N_t^a$ and to split the trajectory into excursions
between encounters (such three excursions are drawn by different
colors).}
\label{fig:disk_traj}
% load('BM5.mat');  A_localtime7_fig_simu(X,Y,In);
\end{figure}

When a Brownian particle diffuses in a geometric confinement, its
encounters with the reflecting boundary can be characterized by the
boundary local time $\ell_t$, which plays the central role in the
theory of stochastic processes \cite{Levy,Ito,Freidlin}.  In the basic
setting of ordinary diffusion, one considers reflected Brownian motion
$\X_t$, released at time $t = 0$ from a fixed point $\x_0$ and
diffusing inside an Euclidean domain $\Omega \subset \R^d$ with
diffusion coefficient $D$ and normal reflections on the smooth
boundary $\pa$.  The boundary local time $\ell_t$ of this process on a
subset of the boundary, $\Gamma \subset \pa$, is defined as
\begin{equation}  \label{eq:ellt_def}
\ell_t = \lim\limits_{a\to 0} \frac{D}{a} \underbrace{\int\limits_0^t dt' \, \Theta(a - |\X_{t'} - \Gamma|)}_{\textrm{residence time}} ,
\end{equation}
where $\Theta(z)$ is the Heaviside step function, and $|\x - \Gamma|$
denotes the distance between a point $\x$ and the boundary region
$\Gamma$.  In this definition, the integral is the residence time of
reflected Brownian motion $\X_t$ up to time $t$ inside a thin boundary
layer of width $a$ around $\Gamma$: $\Gamma_a = \{ \x\in\Omega ~:~ |\x
- \Gamma|<a \}$ (see
\cite{Darling57,Ray63,Knight63,Agmon84,Berezhkovskii98,Dhar99,Yuste01,Godreche01,Majumdar02,Benichou03,Condamin05,Condamin07,Burov07,Burov11}
and references therein).  In the limit $a\to 0$, the residence time
vanishes but its rescaling by $a$ yields a well-defined nontrivial
limit $\ell_t$.  According to Eq. (\ref{eq:ellt_def}), the boundary
local time $\ell_t$ is a non-decreasing process that remains constant
when $\X_t$ is the bulk, and increases only when $\X_t$ hits the
boundary.
Alternatively, the boundary local time $\ell_t$ can also be written as
\begin{equation}  \label{eq:ellt_Nt}
\ell_t = \lim\limits_{a\to 0} a \, \N_t^a ,
\end{equation}
where $\N_t^a$ is the number of downcrossings of the thin boundary
layer $\Gamma_a$ up to time $t$ (Fig. \ref{fig:disk_traj}), i.e., a
regularized version of the number of encounters of the process with
the region $\Gamma$ (see \cite{Grebenkov20} for further discussion).
For a fixed time $t$, $\ell_t$ is a random variable, which can be
characterized by the probability density $\rho(\ell,t|\x_0)$.  Note
that $\ell_t$ has units of length, whereas $\ell_t/D$ has units of
time per length, reflecting the rescaling by $a$ in
Eq. (\ref{eq:ellt_def}).

While general stochastic properties of the boundary local time were
thoroughly investigated in the past (see
\cite{Saisho87,Papanicolaou90,Bass08} and references therein), its
geometry-specific properties are less known.  For instance, how does
the distribution of the number of encounters depend on the shape of
the confining domain and evolve with time?  When does this number
exceed a prescribed threshold?  To answer these questions, general
spectral expansions for $\rho(\ell,t|\x_0)$ and for the probability
density $U(\ell,t|\x_0)$ of the first-crossing time $\T_\ell$ of a
prescribed threshold $\ell$ by the process $\ell_t$, were established
in \cite{Grebenkov19b,Grebenkov20}.  These spectral expansions rely on
the eigenmodes of the Dirichlet-to-Neumann operator, which are known
explicitly only for some simple domains \cite{Grebenkov20c}.  In
particular, closed analytical formulas were derived for diffusion in a
half-line,
\begin{eqnarray}  \label{eq:rho_1d}
\rho(\ell,t|x_0) &=& \erf\left(\frac{x_0}{\sqrt{4Dt}}\right) \delta(\ell) + \frac{\exp\bigl(-\frac{(x_0+\ell)^2}{4Dt}\bigr)}{\sqrt{\pi Dt}}  \,, \\
\label{eq:U1_1d}
U(\ell,t|x_0) &=& (\ell+x_0) \frac{e^{-(\ell+x_0)^2/(4Dt)}}{\sqrt{4\pi Dt^3}}  ,
\end{eqnarray}
and in the exterior of a ball of radius $R$:
\begin{eqnarray}  \label{eq:rho_3d}
\rho(\ell,t|\x_0) &=& \left(1 - \frac{R}{|\x_0|} \erfc\left(\frac{|\x_0|-R}{\sqrt{4Dt}}\right) \right) \delta(\ell) \\  \nonumber
&+& \frac{e^{-\ell/R}}{|\x_0|} \left( \erf\left(\frac{|\x_0| - R + \ell}{\sqrt{4Dt}}\right) 
+ \frac{R \, e^{-(|\x_0| - R +\ell)^2/(4Dt)}}{\sqrt{\pi Dt}}   \right)  \,, \\
\label{eq:U1_3d}
U(\ell,t|\x_0) &=& \frac{R \, e^{-\ell/R}}{|\x_0|} \, \frac{|\x_0|-R+\ell}{\sqrt{4\pi Dt^3}} e^{-(|\x_0| - R +\ell)^2/(4Dt)} \,,
\end{eqnarray}
where $\delta(\ell)$ is the Dirac distribution (see
\cite{Grebenkov20b,Grebenkov20c} for details).  

In turn, the analysis of diffusion in the exterior of a compact planar
domain turns out to be more subtle.  As planar diffusion is recurrent,
the diffusing particle never escapes to infinity and repeatedly
returns to the boundary so that the boundary local time $\ell_t$ grows
to infinity as $t\to \infty$ and thus crosses any threshold with
probability $1$.  However, the probability of not hitting the boundary
up to time $t$ is known to decay {\it logarithmically slowly} with $t$
\cite{Redner}, so that each return to the boundary may take abnormally
long time.  In this paper, we focus on diffusion outside a disk and
derive explicit integral representations for both probability
densities $\rho(\ell,t|\x_0)$ and $U(\ell,t|\x_0)$.  We use then these
presentations to analyze their asymptotic behavior.  We also
characterize the relative contributions of random trajectories with
different exploration sizes.  Finally, we discuss extensions to more
general planar domains, describe some applications in physics and
chemistry, and outline several open questions.

\section{Formal solution}
\label{sec:general}

In this Section, we recall the formal general solution for
$\rho(\ell,t|\x_0)$ and $U(\ell,t|\x_0)$ from
\cite{Grebenkov19b,Grebenkov20}.  The central relation is the expression
for the moment-generating function of the boundary local time
$\ell_t$:
\begin{equation}  \label{eq:ell_ti_exp}
\E_{\x_0} \{ e^{-q\ell_t} \} = S_q(t|\x_0) ,
\end{equation}
where $S_q(t|\x_0)$ is the survival probability, which satisfies the
(backward) diffusion equation
\begin{equation}  \label{eq:Sq_diff}
\partial_t S_q(t|\x_0) = D \Delta S_q(t|\x_0) \qquad (\x_0 \in \Omega), 
\end{equation}
with the initial condition $S_q(0|\x_0) = 1$ and the mixed
Robin-Neumann boundary condition:
\begin{eqnarray}
\left. (\partial_n + q )S_q(t|\x_0)\right|_{\Gamma} & =& 0,   \label{eq:Sq_BC_Robin}\\
\left. \partial_n S_q(t|\x_0) \right|_{\pa\backslash \Gamma} & =& 0 
\end{eqnarray}
(for unbounded domains, the regularity condition $S_q(t|\x_0)\to 1$ as
$|\x_0|\to \infty$ is also imposed).  Here $\Delta$ is the Laplace
operator acting on $\x_0$, $\partial_n$ is the normal derivative at
the boundary oriented outward the domain, and $q \geq 0$ is a
parameter in Eq. (\ref{eq:ell_ti_exp}), which can be related to the
surface reactivity of the subset $\Gamma$ \cite{Grebenkov20}.  On the
other hand, the moment-generating function is defined via the
probability density of $\ell_t$:
\begin{equation}
\E_{\x_0} \{ e^{-q\ell_t} \} = \int\limits_0^\infty d\ell \, e^{-q\ell} \, \rho(\ell,t|\x_0).
\end{equation}
Inverting this Laplace transform, one formally gets
\begin{equation}   \label{eq:rho_ILT}
\rho(\ell,t|\x_0) = \L_{q,\ell}^{-1} \{ S_q(t|\x_0) \} .
\end{equation}

The first-crossing time $\T_{\ell}$ of a given threshold $\ell \geq 0$
by the boundary local time $\ell_t$ is defined as
\begin{equation}
\T_{\ell} = \inf\{ t>0 ~:~ \ell_t > \ell\} .
\end{equation}
As the boundary local time is a non-decreasing process, the cumulative
distribution function of the first-crossing time is determined as
\begin{eqnarray}  \label{eq:Q_ILT}
Q(\ell,t|\x_0) & =& \P_{\x_0}\{ \T_{\ell} < t\} = \P_{\x_0}\{ \ell_t > \ell\} 
 = 1 - \L_{q,\ell}^{-1}\bigl\{ S_q(t|\x_0)/q \bigr\} \,,
\end{eqnarray}
where we used Eq. (\ref{eq:rho_ILT}).  The probability density of the
first-crossing time follows:
\begin{equation}  \label{eq:U_ILT} 
U(\ell,t|\x_0) = \partial_t Q(\ell,t|\x_0) = \L_{q,\ell}^{-1} \left\{ \frac{-\partial_t S_q(t|\x_0)}{q} \right\} .
\end{equation}
Recalling that $H_q(t|\x_0) = -\partial_t S_q(t|\x_0)$ is the
probability density of the first-passage time to a partially reactive
target $\Gamma$ (see, e.g., \cite{Redner}), one can rewrite
Eq. (\ref{eq:U_ILT}) as
\begin{equation}  \label{eq:Hq_U}
H_q(t|\x_0) = \int\limits_0^\infty d\ell \, q \,e^{-q\ell} \, U(\ell,t|\x_0).
\end{equation}
In other words, the probability density $U(\ell,t|\x_0)$ determines
the whole family of the first-passage times described by $H_q(t|\x_0)$
\cite{Grebenkov20}.

Even though the probability densities $\rho(\ell,t|\x_0)$ and
$U(\ell,t|\x_0)$ are formally determined via Eqs. (\ref{eq:rho_ILT},
\ref{eq:U_ILT}), their analysis requires the Laplace transform
inversion, $\L_{q,\ell}^{-1}$, which may be numerically unstable
\cite{Epstein08}.  Moreover, as the parameter $q$ enters through the
Robin boundary condition (\ref{eq:Sq_BC_Robin}), the dependence of the
survival probability $S_q(t|\x_0)$ on $q$ is {\it implicit} that
renders its analysis even more difficult.  The aim of the present
paper is to derive fully explicit representations for both densities
in the case of diffusion outside a disk and then to use them for the
asymptotic analysis.

\section{Diffusion outside a disk}

In this section, we consider reflected Brownian motion in the exterior
of the disk of radius $R$, $\Omega = \{ \x\in\R^2 ~:~ |\x| > R\}$, and
focus on the boundary local time $\ell_t$ on the boundary of that
disk: $\Gamma = \pa$ (Fig. \ref{fig:disk_traj}).  The rotational
invariance of the domain implies that the probability densities and
related quantities do not depend on the angular coordinate.  In the
following, we replace the starting point $\x_0$ by its radial
coordinate $r_0 = |\x_0|$.

We start by recalling the integral representation for the survival
probability $S_q(t|r_0)$, which can be derived from the classical
solution of a similar heat conduction problem
\cite{Carslaw,Grebenkov18b}
\begin{eqnarray}  \label{eq:Spi}
\fl \qquad
S_q(t|r_0) &=& \frac{2q R}{\pi} \int\limits_0^\infty \frac{dz}{z} \, e^{-z^2 Dt /R^2} \\  \nonumber
\fl \qquad
& \times& \frac{Y_0(z r_0/R) (z J_1(z) + qR J_0(z)) - J_0(z r_0/R) (z Y_1(z) + qR Y_0(z))}
{(z J_1(z)+qR J_0(z))^2 + (z Y_1(z) + qR Y_0(z))^2} \,,
\end{eqnarray}
where $J_\nu(z)$ and $Y_\nu(z)$ are the Bessel functions of the first
and second kind, respectively.  To render the dependence of this
expression on $q$ even more explicit, we represent it as
\begin{equation}  \label{eq:Sq_2d_auxil2}
S_q(t|r_0) = \frac{4}{\pi} \int\limits_0^\infty \frac{dz}{z} e^{-z^2 Dt/R^2}  
\Re \left(\frac{q A(z,r_0/R)}{B(z)/R +q}\right),
\end{equation}
where 
\begin{eqnarray}
A(z,r) & =& \frac{i}{2} \, \frac{(J_0(z) + iY_0(z))(J_0(zr) - i Y_0(zr))}{J_0^2(z) + Y_0^2(z)} \,, \\
B(z) & =& \frac{z(J_0(z) J_1(z) + Y_0(z) Y_1(z)) + i \frac{2}{\pi}}{J_0^2(z) + Y_0^2(z)} \,.
\end{eqnarray}

Substituting Eq. (\ref{eq:Sq_2d_auxil2}) into Eq. (\ref{eq:rho_ILT})
and evaluating the inverse Laplace transform with respect to $q$, we
get
\begin{eqnarray}  \label{eq:rho1_2d}
\rho(\ell,t|r_0) &=&  S_{\infty}(t|r_0) \delta(\ell)  \\   \nonumber
& -&  \frac{4}{\pi R}\int\limits_0^\infty \frac{dz}{z} e^{-z^2 Dt/R^2} \,
\Re \biggl(A(z,r_0/R) B(z) e^{-B(z)\ell/R}\biggr) ,
\end{eqnarray}
where 
\begin{equation}  \label{eq:Sinf_2d_auxil}
S_\infty(t|r_0) = \frac{4}{\pi} \int\limits_0^\infty \frac{dz}{z} e^{-z^2 Dt/R^2} \, \Re \bigl(A(z,r_0/R)\bigr)
\end{equation}
is obtained as the limit of Eq. (\ref{eq:Sq_2d_auxil2}) when $q\to
\infty$.  As in Eqs. (\ref{eq:rho_1d}, \ref{eq:rho_3d}), the first
term in Eq. (\ref{eq:rho1_2d}) accounts for trajectories that did not
hit the boundary up to time $t$, for which the boundary local time
$\ell_t$ remained $0$.
The positive-order moments of the boundary local time are analyzed in
\ref{sec:mean_BLT}.  Similarly, Eq. (\ref{eq:Q_ILT}) implies that the
inverse Laplace transform of $S_q(t|r_0)/q$ with respect to $q$ yields
\begin{equation}  \label{eq:Q1_2d}
Q(\ell,t|r_0) = 1 - \frac{4}{\pi} \int\limits_0^\infty \frac{dz}{z} e^{-z^2 Dt/R^2} \,
 \Re \left(A(z,r_0/R) e^{-B(z) \ell/R} \right).
\end{equation}
The time derivative of this expression gives the probability density
$U(\ell,t|r_0)$ of the first-crossing time:
\begin{equation}  \label{eq:U1_2d}
U(\ell,t|r_0) = \frac{4D}{\pi R^2} \int\limits_0^\infty  dz \, z\, e^{-z^2 Dt/R^2} \,
 \Re \left(A(z,r_0/R) e^{-B(z)\ell/R} \right).
\end{equation}
Note also that setting $\ell = 0$ in Eq. (\ref{eq:U1_2d}) yields
\begin{eqnarray}  \label{eq:U1_2d_ell0}
U(0,t|r_0) & =& \frac{4D}{\pi R^2} \int\limits_0^\infty  dz \, z\, e^{-z^2 Dt/R^2} \,
 \Re \bigl(A(z,r_0/R) \bigr) = H_{\infty}(t|r_0),
\end{eqnarray}
where we used Eq. (\ref{eq:Sinf_2d_auxil}) and thus retrieved the
probability density of the first-passage time to the disk.

The explicit integral representations (\ref{eq:rho1_2d},
\ref{eq:Q1_2d}, \ref{eq:U1_2d}) are the main analytical results of the
paper.  Thanks to the exponential factor $e^{-z^2 Dt/R^2}$, these
integrals rapidly converge for large $z$.  In turn, the integrals in
Eqs. (\ref{eq:rho1_2d}, \ref{eq:Q1_2d}) exhibit logarithmically slow
convergence at small $z$.  A practical solution of this issue is
discussed in \ref{sec:numerics}.

The boundary local time $\ell_t$ remains zero until the first
encounter with the boundary.  As a consequence, the first-crossing
time $\T_\ell$ can be decomposed into two contributions: the
first-passage time from $\x_0$ to the circle, $\T_{0,r_0} = \inf\{ t >
0 ~:~ \ell_t > 0 ~|~ |\X_0| = r_0 \}$, and the first-crossing time of
the level $\ell$ after starting from the circle, $\T_{\ell,R} = \inf\{
t > 0 ~:~ \ell_t > \ell ~|~ |\X_0| = R\}$:
\begin{equation}  \label{eq:decomp}
\T_\ell = \T_{0,r_0} + \T_{\ell,R} .
\end{equation}
The strong Markovian character of reflected Brownian motion and of the
boundary local time, as well as the rotational invariance of the
problem imply that these two contributions are independent variables.
As a consequence, the probability density of $\T_\ell$ can be obtained
by convolving the densities of $\T_{0,r_0}$ and $\T_{\ell,R}$.  The
probability density of the first-passage time to the circle,
$H_\infty(t|r_0)$, has been studied long ago (see
\cite{Redner,Levitz08,Grebenkov18b} and references therein).  As a
consequence, one can focus on the probability density $U(\ell,t|R)$ of
the second contribution $\T_{\ell,R}$, for which $A(z,1) = i/2$, and
Eq. (\ref{eq:U1_2d}) becomes
\begin{equation}  \label{eq:U1R_2d}
U(\ell,t|R) = - \frac{2D}{\pi R^2} \int\limits_0^\infty  dz \, z\, e^{-z^2 Dt/R^2} 
\, \Im \left( e^{-B(z)\ell/R} \right).
\end{equation}
In the next section, we discuss the asymptotic behavior of the
density $U(\ell,t|r_0)$.

\section{Asymptotic analysis}
\label{sec:asympt}

To investigate the asymptotic behavior of the probability density
$U(\ell,t|r_0)$, it is convenient to change the integration variable
in Eq. (\ref{eq:U1_2d}) as
\begin{eqnarray}  \label{eq:U1_2d_auxil}
U(\ell,t|r_0) & =& \frac{4}{\pi t} \int\limits_0^\infty  dz \, z\, e^{-z^2} \,
\Re \biggl(A\bigl(zR/\sqrt{Dt},r_0/R\bigr) e^{-B(zR/\sqrt{Dt})\, \ell/R} \biggr).
\end{eqnarray}

\subsection{Short-time asymptotic behavior}

The limit $t\to 0$ corresponds to the large-$z$ expansions of $A(z,r)$
and $B(z)$:
\begin{eqnarray}
A(z,r) & \simeq& \frac{i}{2\, r^{\frac12}} e^{-iz(r-1)} \qquad (z\to\infty), \\
B(z) & \simeq& \frac{1}{2} + iz + O(1/z) \qquad (z\to\infty).  
\end{eqnarray}
Substituting these approximations into Eq. (\ref{eq:U1_2d_auxil}), we
get
\begin{equation}  \label{eq:U_short}
U(\ell,t|r_0) \simeq \frac{e^{-\ell/(2R)}}{(r_0/R)^{\frac12}} \,  \frac{(r_0-R+\ell) \, e^{ - (r_0-R+\ell)^2/(4Dt)}}{\sqrt{4\pi Dt^3}}  \qquad (t\to 0) .
\end{equation}
Apart from the factor $e^{-\ell/(2R)} \,(r_0/R)^{-\frac12}$, this
expression is identical to Eq. (\ref{eq:U1_1d}) for diffusion in the
half-line, if $x_0 = r_0 -R$ denotes the distance to the boundary.
This is not surprising because the circle looks locally flat at short
times, and the behavior should be close to that of the half-plane.  In
turn, the supplementary factor $e^{-\ell/(2R)}\, (r_0/R)^{-\frac12}$
accounts for the curvature of the boundary.  Note that in the case
$r_0 = R$, the same result could alternatively be derived by analyzing
Eq. (\ref{eq:Qp}) in the short-time limit (i.e., $p\to
\infty$).

In the limit $t\to 0$, the particle does not spend enough time near
the boundary to allow for the boundary local time $\ell_t$ to cross a
given threshold $\ell$, and the probability density of too short
first-crossing times, $t \ll \ell^2/(4D)$, is extremely small, even if
the particle starts {\it on} the boundary.  At first thought, this
statement may sound to contradict the well-known property that
(reflected) Brownian motion that crossed a smooth boundary returns
infinitely many times to that boundary during an infinitely short
period \cite{Morters}.  This seeming contradiction reflects the
crucial difference between the boundary local time, $\ell_t$, and the
number of encounters $N_t^a$ with a thin boundary layer of width $a$
(i.e., the number of returns up to $t$).  The latter is defined for
any $a > 0$ but diverges in the limit $a\to 0$.  This divergence,
$N_t^a \to N_t^0 = +\infty$, is evoked in the above statement about
infinitely many returns.  In turn, the boundary local time $\ell_t$,
which is obtained by rescaling $N_t^a$ by $a$ in
Eq. (\ref{eq:ellt_Nt}), remains a meaningful characteristics of the
encounters with the boundary in the limit $a\to 0$.

\subsection{Long-time asymptotic behavior}

The large-$t$ limit corresponds to the small-$z$ behavior of $A(z,r)$
and $B(z)$:
\begin{eqnarray}  \label{eq:Az_small}
A(z,r) &\simeq& \frac{i}{2}\bigl(1 - B(z) \ln r \bigr)  \qquad (z\to 0), \\
\label{eq:Bz_small} 
B(z) &\simeq& \frac{1}{\ln (1/z) + \ln 2 -\gamma - \frac{\pi i}{2}} \qquad (z\to 0)\,,
\end{eqnarray}
where $\gamma \approx 0.5772\ldots$ is the Euler constant.  We
substitute these expressions in Eq. (\ref{eq:U1_2d_auxil}) and then
neglect $\ln z$ as a slowly varying function to get
\begin{eqnarray}  \nonumber
\fl \qquad
U(\ell,t|r_0) & \simeq& \frac{1}{\pi t} \exp\left(-\frac{(\ell/R) b_t}{b_t^2 + (\pi/2)^2} \right)  
\left[\sin \left(\frac{(\ell/R) \pi/2}{b_t^2 + (\pi/2)^2}\right) \left(1 - \frac{b_t \ln (r_0/R)}{b_t^2 + (\pi/2)^2}\right) \right. \\   
\label{eq:U1_2d_asympt}
\fl \qquad
& + & \left. \cos\left(\frac{(\ell/R) \pi/2}{b_t^2 + (\pi/2)^2}\right) \frac{(\pi/2) \ln (r_0/R)}{b_t^2+(\pi/2)^2}\right],
\end{eqnarray}
where the remaining integral over $z$ gave the factor $1/2$, and we
set $b_t = \ln(\sqrt{4Dt}/R) - \gamma$.  When $b_t \gg \pi/2$ and $b_t
\gg \sqrt{\ell/R}$, we finally get
\begin{equation}  \label{eq:U1_2d_asympt0}
U(\ell,t|r_0) \simeq 2 \frac{\ell/R + \ln(r_0/R)}{t \, [\ln( Dt/R^2)]^2}  \qquad (t\to\infty).
\end{equation}

For comparison, we also present the long-time asymptotic behavior of
the probability density $H_q(t|r_0)$ of the first-passage time to a
partially reactive target, which was derived in \cite{Grebenkov18b}:
\begin{equation}  \label{eq:Hpi_long}
H_q(t|r_0) \simeq \frac{2\biggl(\frac{1}{q R} + \ln(r_0/R)\biggr)}
{t \biggl(\pi^2 + \left[\ln(R^2/(4Dt)) + 2\gamma - \frac{2}{qR}\right]^2\biggr)} \,.
\end{equation}
In the very long time limit, one gets
\begin{equation}  \label{eq:Hpi_long2}
H_q(t|r_0) \simeq 2\frac{\frac{1}{q R} + \ln(r_0/R)}
{t \bigl[\ln(Dt/R^2)\bigr]^2} \,.
\end{equation}
In the particular case $q = \infty$, this behavior coincides with
Eq. (\ref{eq:U1_2d_asympt0}) at $\ell =0$, in agreement with
Eq. (\ref{eq:U1_2d_ell0}).  More generally, both densities
$U(\ell,t|r_0)$ and $H_q(t|r_0)$ that are related by
Eq. (\ref{eq:Hq_U}), exhibit the same dependence on time in the limit
$t\to 0$.  Curiously, the relation (\ref{eq:Hq_U}) is satisfied even
for the asymptotic forms in Eqs. (\ref{eq:U1_2d_asympt0},
\ref{eq:Hpi_long2}).

One can see that the probability density $U(\ell,t|r_0)$ exhibits a
very heavy tail at large $t$.  In particular, any positive moment of
the first-crossing time, $\E_{\x_0}\{ \T_\ell^k\}$, with $k \in \R_+$,
is infinite.  This divergence is caused by the contribution of very
long trajectories that explore the plane and can move far away from
the target but unavoidably return to it due to the recurrent character
of planar diffusion.  The same mechanism leads to the infinite mean
first-crossing time for diffusion in the half-line.  However, the
$t^{-3/2}$ decay of $U(\ell,t|x_0)$ in Eq. (\ref{eq:U1_1d}) ensures
that at least the moments $\E_{x_0}\{ \T_\ell^k\}$ with $k < 1/2$
exist in the one-dimensional setting.

\begin{figure}
\begin{center}
\includegraphics[width=88mm]{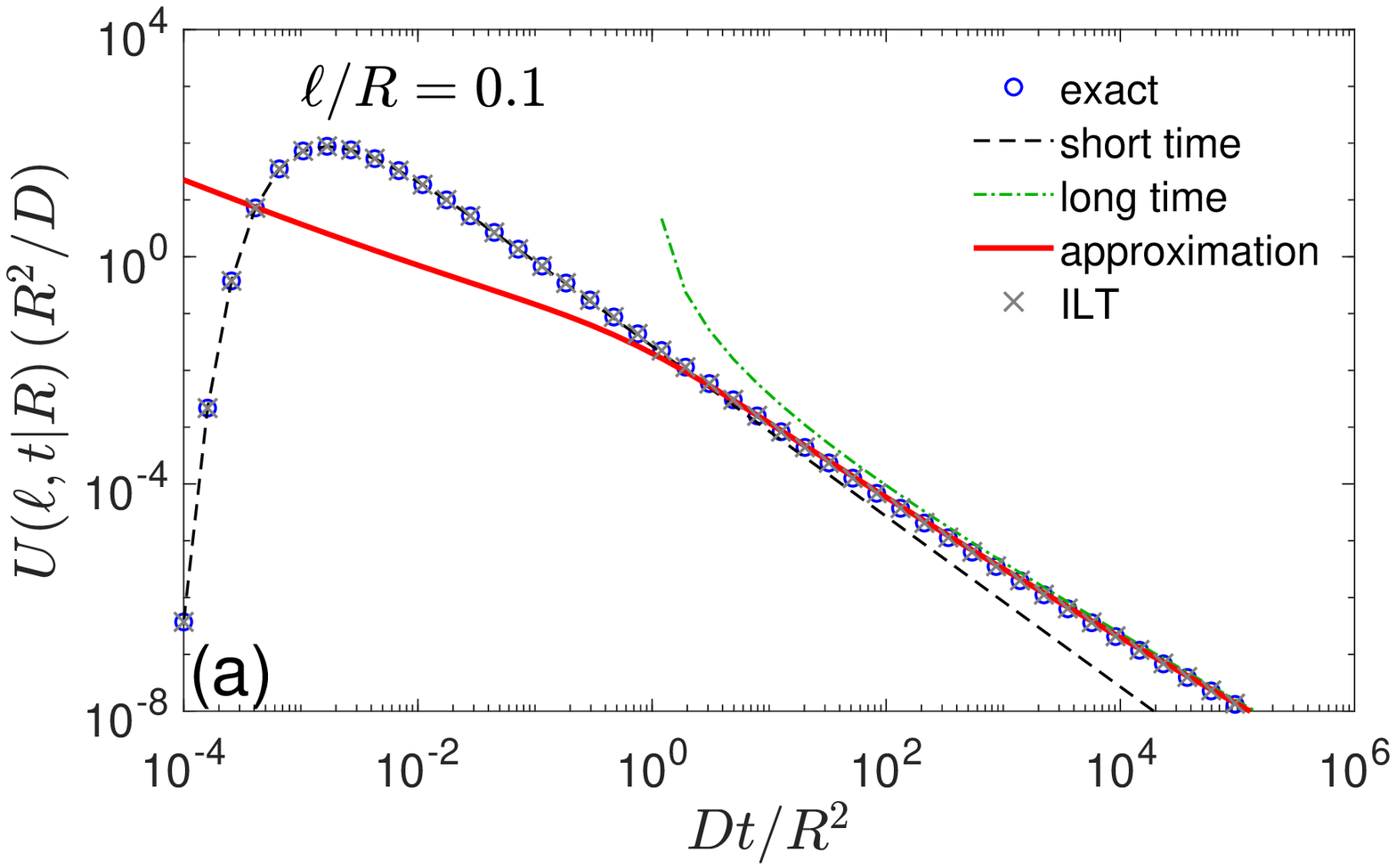} % U_ell01.eps}
\includegraphics[width=88mm]{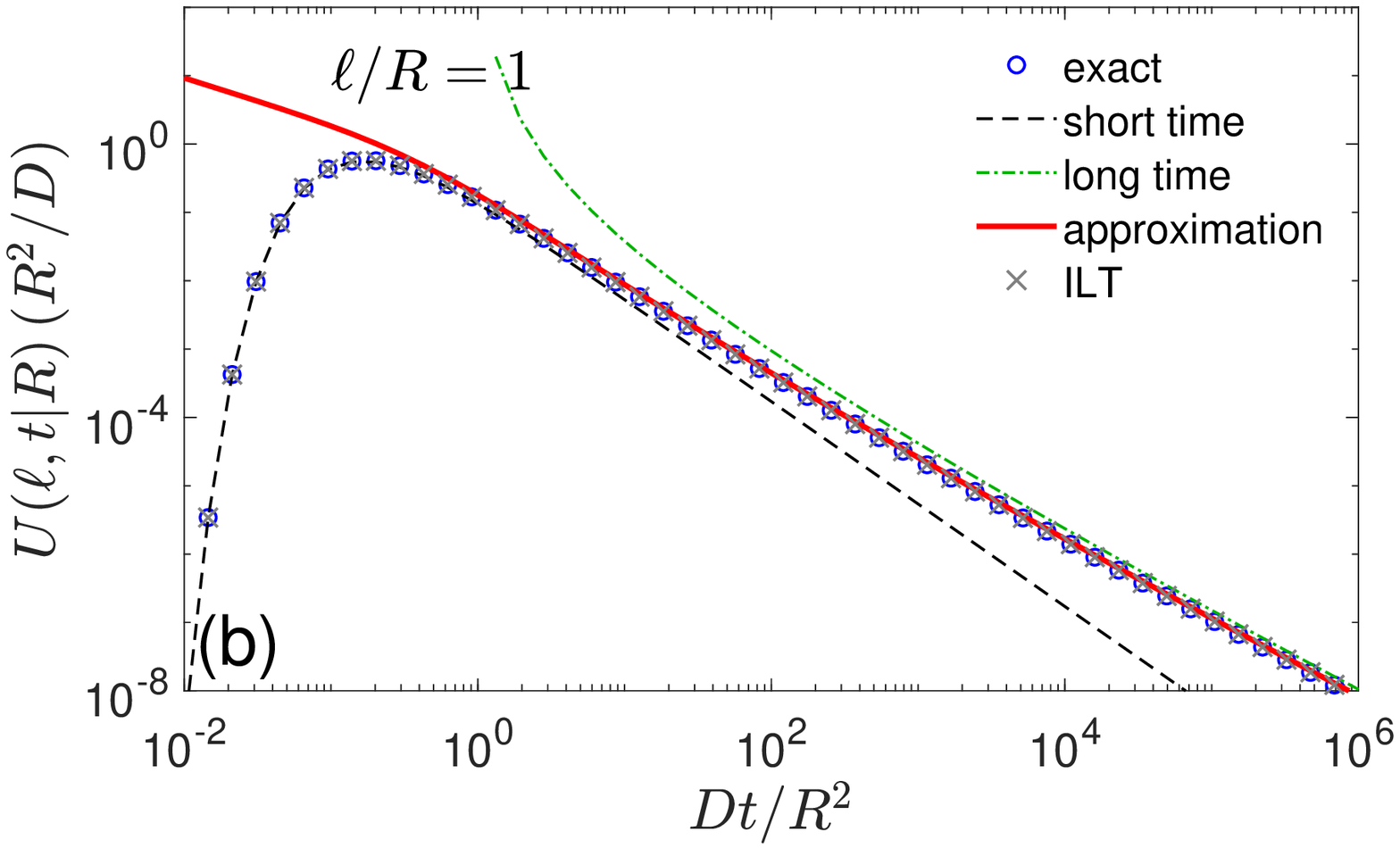} % U_ell1.eps}
\includegraphics[width=88mm]{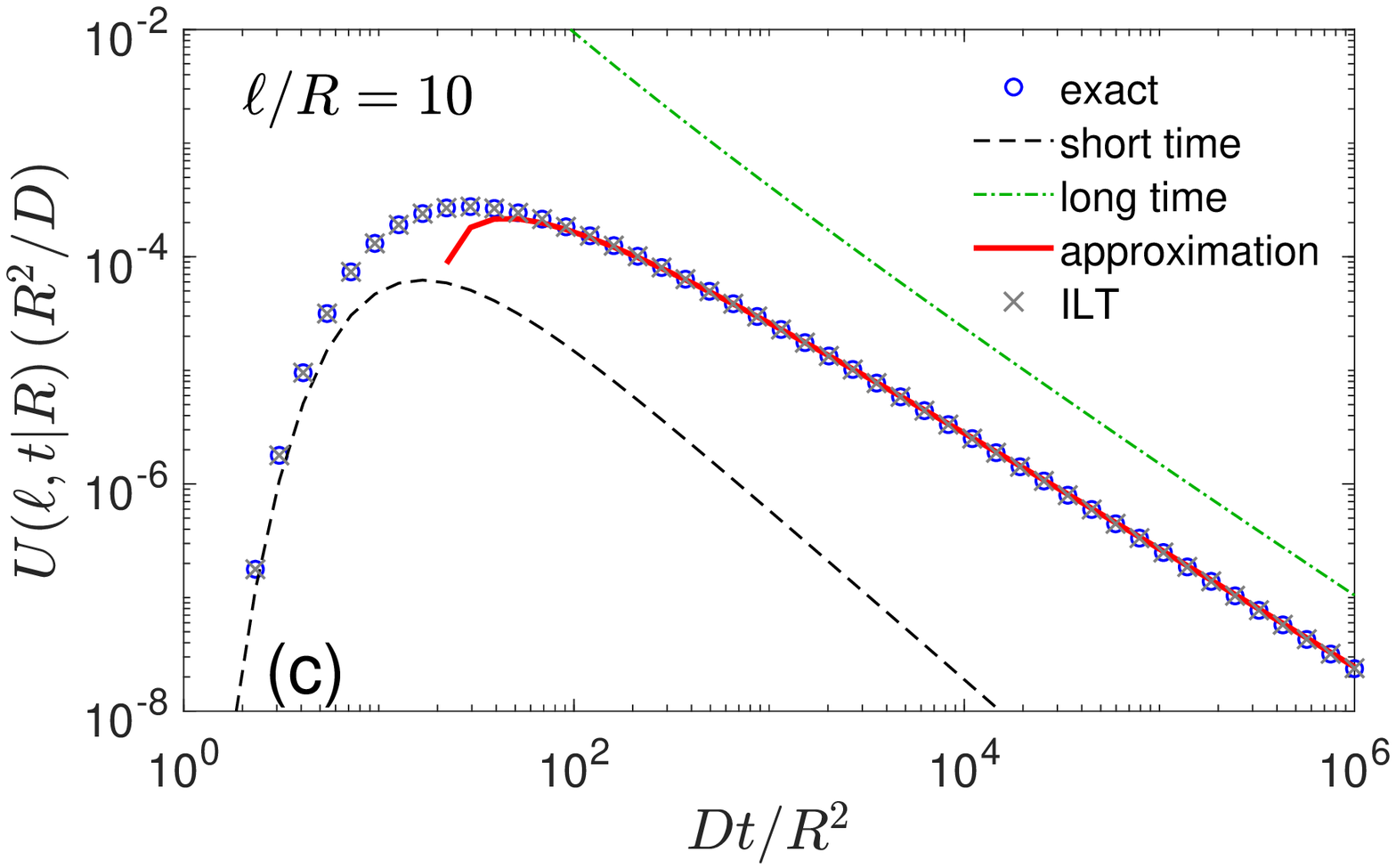} % U_ell10.eps}
\end{center}
\caption{
The probability density $U(\ell,t|R)$ as a function of $t$ for
diffusion outside the disk of radius $R = 1$, with $D = 1$, $\ell =
0.1$ {\bf (a)}, $\ell = 1$ {\bf (b)}, and $\ell = 10$ {\bf (c)}.
Empty circles show the exact solution obtained by numerical
computation of the integral in Eq. (\ref{eq:U1R_2d}); dashed line
presents the short-time asymptotic behavior (\ref{eq:U_short}); solid
and dash-dotted lines show the long-time asymptotic relations
(\ref{eq:U1_2d_asympt}) and (\ref{eq:U1_2d_asympt0}), respectively;
crosses present the numerical computation of the inverse Laplace
transform in Eq. (\ref{eq:U1R_ILT}) by the Talbot algorithm. }
\label{fig:U1_t}
% A_localtime6_U1_2d_fig3c(1);
% A_localtime6_U1_2d_fig3c(2);
% A_localtime6_U1_2d_fig3c(3);
\end{figure}

Figure \ref{fig:U1_t} illustrates the behavior of the probability
density $U(\ell,t|R)$ for several values of the threshold $\ell$.  For
cross-validation, this density was computed by two independent
methods: (i) the numerical evaluation of the integral in
Eq. (\ref{eq:U1R_2d}), and (ii) the numerical computation of the
inverse Laplace transform in Eq. (\ref{eq:U1R_ILT}) by the Talbot
algorithm.  Both methods show an excellent agreement (circles versus
crosses).  At $\ell/R = 0.1$ and $\ell/R = 1$, the short-time
asymptotic relation (\ref{eq:U_short}) accurately captures the
behavior of $U(\ell,t|R)$ at short and even moderate times, $t
\lesssim R^2/D$, but fails at long times, as expected.  In particular,
one can use the simple form of Eq. (\ref{eq:U_short}) to estimate the
most probable first-crossing time: $t_{\rm mp} \simeq (r_0 - R +
\ell)^2/(6D)$.  As the mean value is infinite, the most probable time
plays the role of a natural time scale.  At long times, $t \gg R^2/D$,
Eq. (\ref{eq:U1_2d_asympt0}) correctly captures the long-time behavior
but approaches it slowly, particularly for large $\ell$.  In contrast,
the approximate relation (\ref{eq:U1_2d_asympt}) turns out to be
surprisingly accurate for large and even intermediate times.

The quality of both asymptotic relations (\ref{eq:U_short},
\ref{eq:U1_2d_asympt0}) is much lower in the case $\ell/R = 10$
(Fig. \ref{fig:U1_t}(c)).  On one hand, Eq. (\ref{eq:U_short}) does
not reproduce correctly the maximum of $U(\ell,t|R)$ and captures only
the steep decay at small $t$.  In other words, in the range of
validity of this relation (i.e., $t \lesssim R^2/D$), the density
$U(\ell,t|R)$ is already very small and thus not much useful in
practice.  On the other hand, the approach to
Eq. (\ref{eq:U1_2d_asympt0}) is very slow so that this asymptotic
relation becomes accurate only at extremely large times.  In contrast,
one can still rely on the approximate Eq. (\ref{eq:U1_2d_asympt}).

\subsection{Exploration size of trajectories}

In order to quantify the relative contributions of trajectories with
different spatial extents into the heavy tail of the probability
density $U(\ell,t|R)$, we introduce an absorbing circle of radius $L >
R$ and compute the probability density of the first-crossing time
$\T_\ell$ for the trajectories that remained inside that circle up to
$\T_\ell$.  In order words, while computing the statistics of the
first-crossing times, we discard all trajectories that hit the circle
of radius $L$ up to time $\T_\ell$.  In \ref{sec:annulus}, we describe
an extension of the spectral approach from
\cite{Grebenkov19b,Grebenkov20,Grebenkov20c,Grebenkov20b} to compute
the density $U_L(\ell,t|r_0)$ in the presence of the absorbing circle.
Expectedly, the density $U_L(\ell,t|r_0)$ is not normalized to $1$,
and $1 - \int\nolimits_0^\infty dt \, U_L(\ell,t|r_0)$ is the
probability that the trajectory $\X_t$ has crossed the circle of
radius $L$ before the associated boundary local time could reach the
level $\ell$.  In this setting, we could not derive a fully explicit
representation for $U_L(\ell,t|r_0)$ and thus performed the Laplace
transform inversion in Eq. (\ref{eq:UL_ILT}) numerically by the Talbot
algorithm.

\begin{figure}
\begin{center}
\includegraphics[width=88mm]{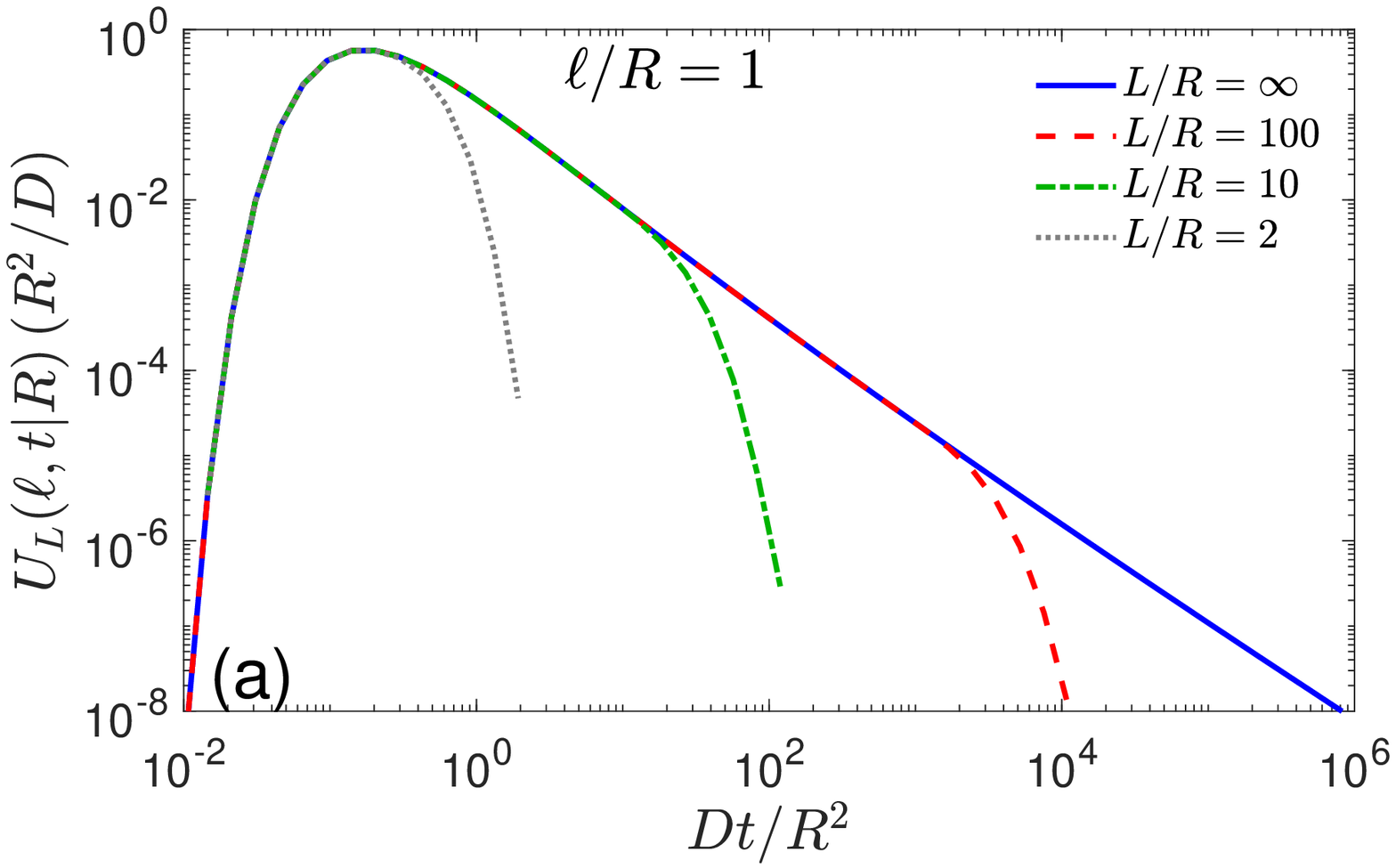} % U_L_ell1.eps}
\includegraphics[width=88mm]{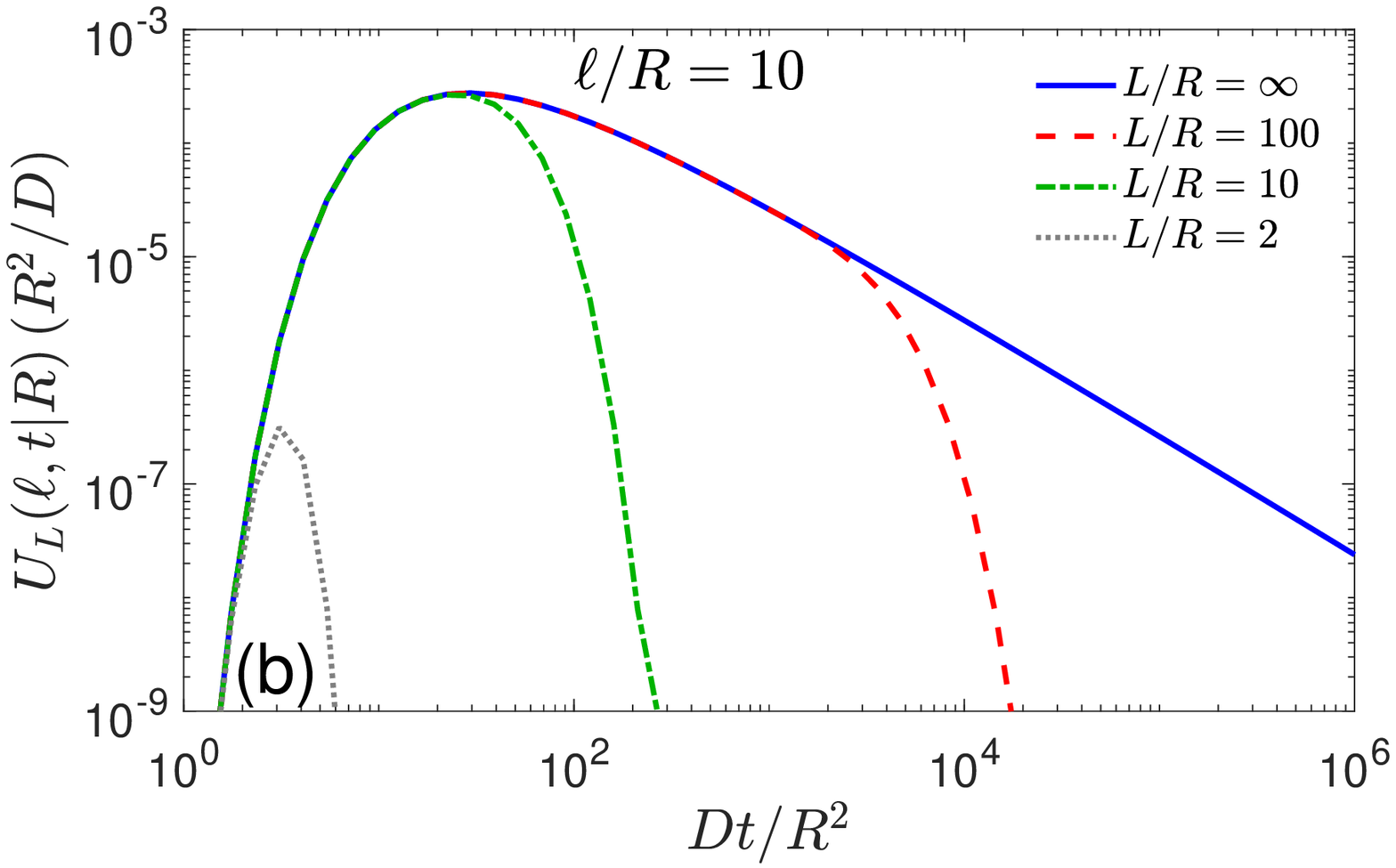} % U_L_ell10.eps}
\end{center}
\caption{
The probability density $U(\ell,t|R)$ (solid line) as a function of
$t$ for diffusion outside the disk of radius $R = 1$, with $D = 1$,
$\ell = 1$ {\bf (a)} and $\ell = 10$ {\bf (b)}.  This density is
compared to the probability densities $U_L(\ell,t|R)$ in the presence
of the absorbing circle of radius $L$, with three values of $L$ as
indicated in the plot.  All these densities were obtained via the
numerical computation of the inverse Laplace transform in
Eq. (\ref{eq:UL_ILT}) by the Talbot algorithm.  Note that two last
curves in the panel {\bf (a)} were truncated at large times to avoid
numerical instabilities of the Laplace transform inversion.}
\label{fig:U1_L}
% A_localtime6_U1_2d_fig4();
\end{figure}

Figure \ref{fig:U1_L} shows the probability densities $U_L(\ell,t|R)$
for several values of $L$ (with $L = \infty$ corresponding to the
former case without absorbing circle).  As the ``restricted''
trajectories explore a bounded domain between two concentric circles
(of radii $R$ and $L$), the particle returns to the inner circle
(target) more often, and the boundary local time increases faster.  As
a consequence, the probability density $U_L(\ell,t|r_0)$ of the
associated first-crossing time is expected to decay much faster as
$t\to\infty$, as confirmed by Fig. \ref{fig:U1_L}.  This faster decay
is also related to the fact that keeping long trajectory within a
bounded region becomes more and more unlikely as time goes on.
This figure explicitly illustrates that the heavy tail of the density
$U(\ell,t|R)$ in Eq. (\ref{eq:U1_2d_asympt0}) is caused by very long
trajectories that explore the unbounded planar space and thus create
long stalling periods, during which the boundary local time does not
change.

\section{Discussion and conclusion}

In this paper, we considered diffusion in the exterior of a disk and
derived the explicit integral representations (\ref{eq:rho1_2d}) and
(\ref{eq:U1_2d}) for the probability densities $\rho(\ell,t|r_0)$ and
$U(\ell,t|r_0)$ of the boundary local time $\ell_t$ and of the
first-crossing time $\T_\ell$, respectively.  As the former density
$\rho(\ell,t|r_0)$ was investigated in \cite{Grebenkov19b}, we mainly
focused on the latter one.  With the aid of the integral
representation (\ref{eq:U1_2d}), we deduced Eqs. (\ref{eq:U_short},
\ref{eq:U1_2d_asympt0}) describing the short-time and the long-time
asymptotic behavior of $U(\ell,t|r_0)$, respectively.  Moreover,
Eq. (\ref{eq:U1_2d_asympt}) was checked to be an accurate
approximation of $U(\ell,t|r_0)$ for moderate and long times.  We
illustrated that the probability density $U(\ell,t|r_0)$ vanishes at
short times, reaches the maximum at intermediate times $\sim
\frac{1}{D} (r_0-R+\ell)^2$, and exhibits a heavy long-time tail such
that all positive-order moments are infinite.  To quantify the effect
of long excursions between encounters, we also analyzed the
probability density $U_L(\ell,t|r_0)$ in the presence of an absorbing
circle of radius $L$.

As discussed in Secs. \ref{sec:intro} and \ref{sec:general}, the
boundary local time $\ell_t$ is a proxy of the number of encounters,
$\N_t^a$, of a diffusing particle with the boundary layer $\Gamma_a$
of width $a$ up to time $t$ (here, $\Gamma = \pa$).  Consequently, the
first-crossing time $\T_\ell$ describes the moment when the number of
encounters exceeds a prescribe threshold.  While the random variables
$\ell_t$ and $\T_\ell$ characterize the reflecting boundary, they form
a natural ground for incorporating surface reactivity and to describe
various diffusion-mediated surface phenomena such as
diffusion-influenced chemical reactions, permeation across biological
membranes, surface relaxation in nuclear magnetic resonance,
etc. \cite{Grebenkov07a,Grebenkov20}.
For instance, $U(0,t|r_0) = H_\infty(t|r_0)$ describes the
first-passage time to the perfectly reactive boundary
\cite{Smoluchowski17}, whereas $H_q(t|r_0)$, expressed through 
Eq. (\ref{eq:Hq_U}) in terms of $U(\ell,t|r_0)$, determines the
reaction time on a partially reactive boundary
\cite{Collins49,Sano79,Shoup82,Zwanzig90,Sapoval94,Filoche99,Benichou00,Grebenkov03,Berezhkovskii04,Grebenkov06,Grebenkov06a,Bressloff08,Reingruber09,Lawley15,Grebenkov17,Bernoff18b,Grebenkov19}.
More sophisticated surface reaction mechanisms can also be implemented
with the help of $U(\ell,t|r_0)$, as discussed in \cite{Grebenkov20}.

Interestingly, the probability densities $\rho(\ell,t|r_0)$ and
$U(\ell,t|r_0)$ also describe the first-encounter statistics of two
independent particles diffusing in the plane.  In fact, associating
the coordinate frame with one of the particles, the diffusive dynamics
of two circular particles of radii $R_1$ and $R_2$ with diffusivities
$D_1$ and $D_2$ can be mapped onto planar diffusion of a single
point-like particle with diffusivity $D = D_1 + D_2$ toward a static
circular target of radius $R = R_1 + R_2$.  As a consequence,
$\ell_t/a$ is the number of encounters of such two particles, where
$a$ is the separation between two particles at each encounter.  In
other words, a single encounter is counted when two particles hit each
other and then moved away and become separated by a distance exceeding
$a$.  In turn, $\T_{aN}$ is the first moment when the number of such
encounters exceeds $N$.  As an encounter of two species is the
necessary step for most bimolecular reactions, reproduction in
animals, and virus spreading in humans, both quantities can be used in
describing various encounters in chemistry, biology and ecology (see
\cite{Tejedor11,Amitai12,Bressloff13,Giuggioli13,Tzou14,Agliari14,Agliari16,Peng19,LeVot20}
and references therein).

Even though the explicit representations (\ref{eq:rho1_2d}) and
(\ref{eq:U1_2d}) were derived for the particular case of a disk, the
asymptotic behavior of the probability density $U(\ell,t|\x_0)$ is
expected to remain valid for diffusion in the exterior of a general
compact planar domain with smooth boundary.  Indeed, at short times,
$t\ll R_c^2/D$, the boundary looks locally flat with respect to the
radius of curvature $R_c$ of the boundary in a vicinity of the
starting point.  As a consequence, the time dependence of the density
$U(\ell,t|\x_0)$ is expected to be given by Eq. (\ref{eq:U1_1d}) for
the one-dimensional setting, where $x_0 = |\x_0 - \Gamma|$ is the
distance between the starting point $\x_0$ and the target $\Gamma$.
In turn, the correction factor $e^{-\ell/(2R)} \, (r_0/R)^{-\frac12}$
from Eq. (\ref{eq:U_short}) is specific to the disk.  Finding the
shape-dependent correction factor for other domains presents an
interesting perspective.  Similarly, at large times, long excursions
between encounters are expected to lead to a heavy tail of the
probability density $U(\ell,t|\x_0)$ as $t\to
\infty$.  The characteristic $1/(t\ln^2(t))$ decay may remain
universal but the dependence on $\ell$ and $\x_0$ could be
shape-specific.  Validation and specification of these conjectures
present another exciting direction for future research.  This problem
is tightly related to the asymptotic behavior of the spectrum of the
Dirichlet-to-Neumann operator associated with the modified Helmholtz
equation \cite{Grebenkov19b} (see also \ref{sec:annulus}).

%\begin{acknowledgments}
\section*{Acknowledgments}

The author acknowledges a partial financial support from the Alexander
von Humboldt Foundation through a Bessel Research Award.
%\end{acknowledgments}

\appendix
\section{Long-time behavior of the mean boundary local time}
\label{sec:mean_BLT}

According to Eq. (\ref{eq:rho1_2d}), the probability density
$\rho(\ell,t|r_0)$ exhibits a rapid decay for large $\ell$, ensuring
the existence of all positive-order moments of the boundary local time
$\ell_t$:
\begin{equation}  \label{eq:ellk_def}
\E_{\x_0}\{ \ell_t^k\} = \int\limits_0^\infty d\ell \, \ell^k \, \rho(\ell,t|\x_0)  \qquad (k > 0).
\end{equation}
The explicit representation (\ref{eq:rho1_2d}) can be used to compute
these moments.  However, the direct exchange of the order of integrals
over $\ell$ in Eq. (\ref{eq:ellk_def}) and over $z$ in
Eq. (\ref{eq:rho1_2d}) is not possible as the resulting integral over
$z$ would be divergent.  To overcome this limitation, one can first
express the time derivative of the moment $\E_{\x_0}\{ \ell_t^k\}$ as
\begin{equation} 
\partial_t \E_{\x_0}\{ \ell_t^k\} = \frac{4 D R^{k-2} \Gamma(k+1)}{\pi} \int\limits_0^\infty dz\, z\, e^{-z^2 Dt/R^2} \,
\Re \left(\frac{A(z,r_0/R)}{[B(z)]^k}\right).
\end{equation}
Rescaling the integral variable yields
\begin{equation} 
\partial_t \E_{\x_0}\{ \ell_t^k\} = \frac{4 R^{k} \Gamma(k+1)}{\pi t} \int\limits_0^\infty dz\, z\, e^{-z^2}  \,
\Re \left(\frac{A(zR/\sqrt{Dt},r_0/R)}{\bigl[B(zR/\sqrt{Dt})\bigr]^k}\right).
\end{equation}

In the long-time limit, one uses the small-$z$ expansions
(\ref{eq:Bz_small}, \ref{eq:Az_small}) to get
\begin{eqnarray}   \nonumber
\partial_t \E_{\x_0}\{ \ell_t^k\} & \simeq& \frac{2R^k \Gamma(k+1)}{\pi t} \int\limits_0^\infty dz \, z \, e^{-z^2} \,
\Re \left(i \biggl[(b_t - \ln z - \pi i/2)^k \right.\\
& -& \left. \ln (r_0/R) (b_t - \ln z - \pi i/2)^{k-1} \biggr]\right),
\end{eqnarray}
where $b_t = \ln(\sqrt{4Dt}/R) - \gamma$.  Integrating this expression
over $t$, we get
\begin{eqnarray}   \nonumber
\E_{\x_0}\{ \ell_t^k\} & \simeq& \frac{2R^k \Gamma(k+1)}{\pi} \int\limits_0^\infty dz \, e^{-z} \,
\Im \left(- \frac{(b_t - \frac12 \ln z - \frac12 \pi i)^{k+1}}{k+1} \right.\\
& +& \left. \ln (r_0/R) \frac{(b_t - \frac12 \ln z - \frac12 \pi i)^k}{k} \right),
\end{eqnarray}
where we also changed the integration variable $z^2 \to z$.  Using the
binomial expansion, one can evaluate these integrals term by term.  In
particular, we get the mean and the variance in the long-time limit:
\begin{eqnarray}
\E_{\x_0}\{ \ell_t\} & \simeq& R \biggl(\ln\bigl(\sqrt{4Dt}/r_0\bigr) - \gamma/2\biggr)  , \\  \nonumber
\var_{\x_0}\{ \ell_t\} & =& \E_{\x_0}\{ \ell_t^2\} - \bigl(\E_{\x_0}\{ \ell_t\}\bigr)^2 \\
& \simeq& R^2 \left( \bigl(\ln\bigl(\sqrt{4Dt}/R\bigr) - \gamma/2\bigr)^2 - \ln^2(r_0/R) - \frac{\pi^2}{12}\right)  .
\end{eqnarray}
These expressions generalize the former results from
\cite{Grebenkov19b} to an arbitrary starting point.

\section{Numerical computation}
\label{sec:numerics}

Using Eqs. (\ref{eq:Bz_small}, \ref{eq:Az_small}), one can easily
check that the integrals in Eqs. (\ref{eq:rho1_2d}, \ref{eq:Q1_2d})
converge logarithmically slowly near $z = 0$.  To improve the accuracy
of numerical computations, one can choose an appropriate $\ve \ll 1$
and split the integral into two parts, from $0$ to $\ve$, and from
$\ve$ to $\infty$.  The second integral is then evaluated numerically,
whereas the contribution of the first integral, $I_\ve$, can be found
approximately.

Let us first consider the computation of $Q(\ell,t|r_0)$ in
Eq. (\ref{eq:Q1_2d}), for which
\begin{eqnarray*}
I_\ve & = &\frac{4}{\pi}\int\limits_0^\ve \frac{dz}{z} e^{-z^2 Dt/R^2} \Re \left(A(z,r) e^{-B(z) \ell/R} \right) \\
&\approx& \frac{4}{\pi}\int\limits_0^\ve \frac{dz}{z} \Re \left(\frac{i}{2} \bigl(1 - B(z)\ln r \bigr) e^{-B(z) \ell/R} \right) \\
& \approx& - \frac{2}{\pi}\int\limits_{\ln (1/\ve)}^\infty dy \, \Im \left(\biggl(1 - \frac{\ln r}{y+b}\biggr) 
\exp\biggl(- \frac{c}{y+b}\biggr) \right) ,
\end{eqnarray*}
where $c = \ell/R$ and $b = \ln 2 - \gamma - \frac{1}{2} \pi i$.  Here
we approximated $e^{-z^2 Dt/R^2} \approx 1$ for $0\leq z\leq\ve$, and
made the change of variables: $y = \ln (1/z)$.  Note that the above
approximation naturally imposes the constraint on $\ve$: $\ve^2
\ll R^2/(Dt)$.  In particular, as $t$ increases, one has to choose
smaller and smaller $\ve$.

The last integral can be evaluated as
\begin{equation}
I_\ve \approx 1 + \frac{2}{\pi} \,\Im \biggl( b_\ve \, e^{-c/b_\ve} - \bigl(c + \ln(r_0/R)\bigr) \, \Ei_1\bigl(c/b_\ve\bigr)\biggr) ,
\end{equation}
where $\Ei_1(z) = \int\nolimits_1^\infty dk \, e^{-zk}/k =
\Gamma(0,z)$ is the exponential integral, and $b_\ve = b +
\ln(1/\ve)$.  As a consequence, we get
\begin{equation}
Q(\ell,t|r_0) = 1 - I_\ve - \frac{4}{\pi}\int\limits_\ve^\infty \frac{dz}{z} e^{-z^2 Dt/R^2}\, 
\Re \left(A(z,r_0/R) e^{-B(z) \ell/R} \right).
\end{equation}

A similar correction can be used for an accurate computation of
$\rho(\ell,t|r_0)$:
\begin{eqnarray} \label{eq:rho1_2d_bis}
\rho(\ell,t|r_0) & =&  S_{\infty}(t|r_0) \delta(\ell) - I_\ve^{(\rho)} \\  \nonumber
& -&  \frac{4}{\pi R}\int\limits_{\ve}^\infty \frac{dz}{z} e^{-z^2 Dt/R^2} \,
\Re \left(A(z,r_0/R) B(z) e^{-B(z)\ell/R}\right) ,
\end{eqnarray}
with
\begin{eqnarray} \nonumber
I_\ve^{(\rho)} &=& \frac{4}{\pi R}\int\limits_0^{\ve} \frac{dz}{z} e^{-z^2 Dt/R^2} \,
\Re \left(A(z,r_0/R) B(z) e^{-B(z)\ell/R}\right) \\
&\approx& \frac{2}{\pi R} \, \Im \left(\Ei_1(c/b_\ve) - \frac{\ln(r_0/R)}{c} e^{-c/b_\ve} \right). 
\end{eqnarray}
Note that there is no need for such a correction for computing
$U(\ell,t|r_0)$ because $I_\ve$ does not depend on time and thus
disappears after differentiating $Q(\ell,t|r_0)$ with respect to $t$.

\section{Alternative representation}
\label{sec:Qp}

In \cite{Grebenkov19b,Grebenkov20}, alternative representations for
the probability densities $\rho(\ell,t|\x_0)$ and $U(\ell,t|\x_0)$
were developed in terms of the eigenmodes of the Dirichlet-to-Neumann
operator $\M_p$ for an arbitrary Euclidean domain $\Omega$ with a
smooth bounded boundary $\pa$.  For a given function $f$ on $\pa$, the
operator $\M_p$ associates another function $g = (\partial_n
w)|_{\pa}$ on that boundary, where $w$ satisfies the modified
Helmholtz equation $(p - D\Delta) w = 0$ in $\Omega$ with Dirichlet
boundary condition $w|_{\pa} = f$
\cite{Arendt14,Daners14,Arendt15,Hassell17,Girouard17}.  For instance,
for diffusion outside the disk of radius $R$, the probability
$Q(\ell,t|R)$ reads
\begin{equation}  \label{eq:Qp}
Q(\ell,t|R) = \L_{p,t}^{-1}\biggl\{ \frac{1}{p} \exp\bigl(- \mu_0^{(p)} \ell \bigr)\biggr\} ,
\end{equation}
where $\L_{p,t}^{-1}$ is the inverse Laplace transform with respect to
$p$, 
\begin{equation}  \label{eq:mu0_disk}
\mu_0^{(p)} = \sqrt{p/D} \frac{K_1(R\sqrt{p/D})}{K_0(R\sqrt{p/D})}
\end{equation}
is the smallest eigenvalue of the Dirichlet-to-Neumann operator $\M_p$
in the exterior of the disk, and $K_\nu(z)$ is the modified Bessel
function of the second kind \cite{Grebenkov19b}.  The densities
$\rho(\ell,t|R)$ and $U(\ell,t|R)$ follow immediately by
taking the derivatives with respect to $\ell$ and $t$, respectively:
\begin{equation}
\rho(\ell,t|R) = \L_{p,t}^{-1} \biggl\{ \frac{\mu_{0}^{(p)}}{p} \exp(-\mu_{0}^{(p)} \ell) \biggr\} .
\end{equation}
and
\begin{equation}  \label{eq:U1R_ILT}
U(\ell,t|R) = \L_{p,t}^{-1} \biggl\{ \exp(-\mu_{0}^{(p)} \ell) \biggr\} .
\end{equation}

In this Appendix, we briefly describe a direct derivation of
Eq. (\ref{eq:U1R_ILT}) that illustrates the meaning of the boundary
local time as a proxy for the number of encounters.  Let us consider a
regularized version of the problem, in which the particle after
hitting the circle at some point $(R,\varphi)$ (in polar coordinates)
is released from a bulk point $(R + a,\varphi)$, a distance $a$ above
the circle (Fig. \ref{fig:disk_traj}).  The first-crossing time
$\T_\ell$ can then be represented as $\T_\ell = \tau_1 + \tau_2 +
\ldots + \tau_{N}$, where $\tau_i$ is the duration of bulk diffusion
after $i$-th release (here we assumed that the particle starts on the
circle), and $N \approx \ell/a$ is the number of such bulk
explorations, see Eq. (\ref{eq:ellt_Nt}).  The random variables
$\tau_i$ are independent and identically distributed, with a
well-known moment-generating function \cite{Redner}:
\begin{equation}
\E\{ e^{-p\tau_i}\} = \frac{K_0\bigl((R+a)\sqrt{p/D}\bigr)}{K_0\bigl(R\sqrt{p/D}\bigr)} \,.
\end{equation}  
As a consequence, one has
\begin{eqnarray*}
\E\{ e^{-p\T_\ell} \} & =& \biggl( \E\{ e^{-p\tau_1}\}\biggr)^{N} \\
& =& \exp\left(N \ln \left(\frac{K_0((R+a)\sqrt{p/D})}{K_0(R\sqrt{p/D})}\right)\right) \\
& \simeq& \exp\left(\frac{\ell}{a} \ln \left(1 + \frac{K'_0(R\sqrt{p/D})}{K_0(R\sqrt{p/D})} a \sqrt{p/D}\right)\right) \\
& \to & \exp\Biggl(- \ell \underbrace{\sqrt{p/D} \frac{K_1(R\sqrt{p/D})}{K_0(R\sqrt{p/D})}}_{=\mu_0^{(p)}} \Biggr) \qquad (a\to 0).
\end{eqnarray*}
As a consequence, the inverse Laplace transform of the right-hand side
with respect to $p$ yields $U(\ell,t|R)$ in Eq. (\ref{eq:U1R_ILT}).

\section{Concentric annulus}
\label{sec:annulus}

In order to quantify the relative contributions of far-reaching
trajectories, it is convenient to introduce an absorbing circle of
radius $L$.  In other words, we are interested in diffusion inside a
circular annulus, $\Omega = \{ \x\in\R^2 ~:~ R < |\x|<L\}$.  In this
Appendix, we adopt the spectral approach from
\cite{Grebenkov19b,Grebenkov20,Grebenkov20b,Grebenkov20c} to deduce
the distribution of the boundary local time on the inner circle with
the constraint of not hitting the outer circle.  For brevity, we skip
details and only sketch the main formulas.

Due to rotational invariance, the eigenfunctions of the
Dirichlet-to-Neumann operator $\M_p$ associated to the inner circle
are the Fourier harmonics,
\begin{equation}
v_n^{(p)} = \frac{e^{in\varphi}}{\sqrt{2\pi R}}  \qquad (n\in \Z),
\end{equation}
whereas the associated eigenvalues are
\begin{equation}   \label{eq:mu0_annulus}
\mu_{n,L}^{(p)} = - \alpha \frac{K_n(\alpha L) I'_n(\alpha R) - I_n(\alpha L) K'_n(\alpha R)}
{K_n(\alpha L) I_n(\alpha R) - I_n(\alpha L) K_n(\alpha R)}  \,,
\end{equation}
with $\alpha = \sqrt{p/D}$ and the prime denoting the derivative with
respect to the argument (see similar computations in
\cite{Grebenkov19b,Grebenkov20c}).  The decomposition
(\ref{eq:decomp}) of the first-crossing time $\T_\ell$ implies that
its probability density is obtained by convolution of two densities,
which in the Laplace domain reads
\begin{equation}  \label{eq:UL_ILT}
U_L(\ell,t|r_0) = \L_{p,t}^{-1} \biggl\{ \tilde{H}_L(p|r_0) \, \exp(-\mu_{0,L}^{(p)} \ell) \biggr\} .
\end{equation}
Here
\begin{equation}
\tilde{H}_L(p|r_0) = \frac{K_0(\alpha L) I_0(\alpha r_0) - I_0(\alpha L) K_0(\alpha r_0)}{K_0(\alpha L) I_0(\alpha R) - I_0(\alpha L) K_0(\alpha R)} 
\end{equation}
is the Laplace transform of the probability density of the
first-passage time to the inner circle in the presence of an outer
absorbing circle (expectedly, this function vanishes on the outer
circle, $r_0 = L$, and is equal to $1$ on the inner circle, $r_0 =
R$).  In turn, $\exp(-\mu_{0,L}^{(p)} \ell)$ is the Laplace transform
of the probability density $U(\ell,t|R)$ (compare with
Eq. (\ref{eq:U1R_ILT})).  In the limit $L\to\infty$, the eigenvalue
$\mu_{0,L}^{(p)}$ from Eq. (\ref{eq:mu0_annulus}) tends to
$\mu_0^{(p)}$ from Eq. (\ref{eq:mu0_disk}), and one retrieves
Eq. (\ref{eq:U1R_ILT}).  Similarly, one gets
\begin{equation}
Q_L(\ell,t|r_0) = \L_{p,t}^{-1} \left\{ \tilde{H}_L(p|r_0) \,  \frac{\exp(-\mu_{0,L}^{(p)} \ell)}{p}\right\} 
\end{equation}
and
\begin{equation}
\rho_L(\ell,t|r_0) = \L_{p,t}^{-1} \left\{ \tilde{H}_L(p|r_0) \, \frac{\mu_{0,L}^{(p)}}{p} \exp(-\mu_{0,L}^{(p)} \ell) \right\} .
\end{equation}

To get the long-time behavior, one can evaluate the above expressions
in the limit $p\to 0$; in particular,
\begin{equation}
\lim\limits_{p\to 0}\tilde{H}_L(p|r_0) = \frac{\ln(L/r_0)}{\ln(L/R)} \,, \qquad  \lim\limits_{p\to 0} \mu_0^{(p)} = \frac{1}{R \ln(L/R)} \,,
\end{equation}
from which
\begin{equation}
Q_L(\ell,\infty|R) = \frac{\ln(L/r_0)}{\ln(L/R)} \exp\biggl(- \frac{\ell/R}{\ln (L/R)}\biggr) .
\end{equation}
This is the probability that the threshold $\ell$ is crossed by the
boundary local time $\ell_t$ {\it before} the trajectory $\X_t$ hit
the outer circle of radius $L$.  As $L\to \infty$, this probability
tends to $1$, but this approach is slow and controlled by $\ln (L/R)$.

\end{document}